\documentstyle[12pt,epsfig]{article}
\def\NPB{{ Nucl. Phys.} B}
\def\PLB{{ Phys. Lett.}  B}
\def\PRL{ Phys. Rev. Lett.}
\def\PRD{{ Phys. Rev.} D}

\def\be{\begin{equation}}
\def\ee{\end{equation}}
\def\bea{\begin{eqnarray}}
\def\eea{\end{eqnarray}}

\def\PRL#1#2#3{{ Phys. Rev. Lett.} { #1}~(#3) #2}
\def\PRD#1#2#3{{ Phys. Rev.} { D#1}~(#3) #2}
\def\PLB#1#2#3{{ Phys. Lett.} { B#1}~(#3) #2}

\def\NPB#1#2#3{{ Nucl. Phys.} { B#1}~(#3) #2}

\def\beq{\begin{equation}}
\def\eeq{\end{equation}}
\def\bea{\begin{eqnarray}}
\def\eea{\end{eqnarray}}

\def\beq{\begin{equation}}
\def\eeq{\end{equation}}
\def\bea{\begin{eqnarray}}
\def\eea{\end{eqnarray}}
\def\ba{\begin{array}}

\def\ea{\end{array}}

\def\bbz{fa Z \kern-8.9pt Z}
  \newcommand{\ccaption}[2]{
    \begin{center}
    \parbox{1.0\textwidth}{
      \caption[#1]{\small{{#2}}}
      }
    \end{center}
    }

\begin{document}


\vspace*{-1cm}
\begin{flushright}
{CERN-TH/97-135}
\end{flushright}

\vspace{0.1cm}
\begin{center}
{\large {\bf 
R-parity violation and the New Events at HERA}} \\

\vspace{.2cm}                                                  
\end{center}

\begin{center}
\vspace{.3cm}
{\large S. Lola} \\
\vspace{.3cm}
{ Theory Division, CERN, CH-1211, Geneva 23, Switzerland}
\end{center}                              

\begin{center}
\noindent
{\em Talk 
at the $5^{th}$ 
  International Workshop ``DIS 97'',
  Chicago, Apr. 1997} \\
{\em Work done with G. Altarelli, J. Ellis, G.F. Giudice and M.L. Mangano}
\end{center}

\vspace{0.3 cm}

{\bf Abstract.} 
{\small We summarise 
possible explanations of the HERA large-$Q^2$ data,
in the framework of R-parity violating supersymmetry.
Experimental limits indicate that
the most likely production channels are
$e^+d\rightarrow \tilde c_L$, 
$e^+d\rightarrow \tilde t$ and
$e^+s\rightarrow
\tilde t$. We study the regions of the parameter space
that lead to consistent branching ratios, with and without
the unification condition for gaugino masses.
Cancellations in the
coupling of the lightest
neutralino to $\tilde{c}_L$,
result in a balance
between R-parity violating and R-parity conserving decay
modes.
Such cancellations are not present
in the coupling of the neutralino with
other particles and an interesting case is
$\tilde{\nu}_{L}$, which
could be produced at LEP2 via an $L_1L_{2,3}\bar{E}_1$ operator.
On the other hand, 
the $\tilde{t}$ branching ratios
depend mainly 
on the mass of the lightest chargino and
tend to be dominated 
by either the R-conserving or the R-violating mode.}

\vspace{0.2 cm}

The experiments H1~\cite{H1} and ZEUS~\cite{ZEUS} 
at HERA have reported an excess of deep-inelastic $e^+p$
scattering events at large values of $Q^2$.
The events of H1 suggest a resonance with
$e^+$-quark quantum
numbers and a mass around
$200$ GeV, while  the ZEUS data points are more scattered in mass.
More data will be needed, in order to clarify whether the excess 
is just a statistical fluctuation
or an indication of new physics.
However, in the meantime,
it is important to pursue different possible 
interpretations of the HERA data~\cite{NEWH}.
Among the various schemes that have been proposed, 
R-parity violating supersymmetry
seems to be a very promising possibility\footnote{
Alternative schemes and their possible effects 
have been discussed at this conference~\cite{DIS97}.}.

The R-parity violating superpotential, also contains the couplings
$L_iL_j{\bar E}_k$, $L_iQ_j{\bar D_k}$ and
${\bar U_i}{\bar D_j}{\bar D_k}$,
where $L (Q)$ are the left-handed lepton (quark) superfields,
while ${\bar
E},{\bar D},$ and ${\bar U}$ are the corresponding right-handed fields.
It is possible as a result of symmetries~\cite{MODELS}, to
allow the violation of only a subset of these operators,
while being consistent with the limits on proton decay.
Among the R-parity violating couplings,
$9$ could in principle lead to resonant squark production
at HERA. However, 
if one requires to match the HERA data, while satisfying
the various experimental 
constraints, not all possibilities survive.
The squark production mechanisms permitted by the $\lambda'$
couplings include $e^+$ and valence $d$ collisions to form
${\tilde u}_L, {\tilde c}_L$ or $\tilde t_L$, 
and collisions with sea quarks, of the
type $e^+ {d_i}$ ($i=2,3$) or $e^+ {\bar u_i}$.
The required magnitude of the coupling 
$\lambda'$ is fixed by the
product of the cross section $\sigma$ and the squark branching ratio 
$\cal B$ for the $R$-parity violating mode ${\tilde q}\to e^+ q'$. 
Then, assuming a total of
10 events in the combined experiments, 
and for the quoted detector
efficiencies, it is found
that for the valence production mechanism 
$\lambda'_{1j1} \approx
0.04/\sqrt{\cal B}$, while for  the sea production
mechanisms $\lambda'>0.3/\sqrt{\cal B}$~\cite{AEGLM}.
Here, we should note that
scalar leptoquarks with
${\cal B}(e^+ q) = 1$,
are strictly bound from Tevatron data, unless their mass
is as high as $210$ GeV~\cite{CDF}. On 
the other hand, a  squark with R-parity violating 
couplings~\cite{pi}, has additional decay modes, 
and accommodates easier the Tevatron constraints.

Among the valence production mechanisms,
$\tilde{u}_L$ production
is ruled out by limits 
from $\beta\beta$ decay~\cite{HirVer}. 
For $\tilde{c}_L$ production, the stricter limit
on $\lambda'_{121}$ arises from 
$K \rightarrow \pi {\bar \nu} \nu$ decays~\cite{Agashe},
$\vert \lambda'_{121} \vert 
< 2 \times 10^{-2} \left({m_{{\tilde d}_{k_R}} / 200 \nonumber
~\hbox{GeV}}\right)$,
thus ${\tilde c}_L$ production at HERA 
implies that 
$m_{{\tilde d}_R}> 400~{\hbox{GeV}}/ \sqrt{\cal B}$.
However, this bound on $m_{{\tilde d}_R}$,
which depends on 
the mixing in the down sector,
can be partially relaxed if
various non-vanishing coupling constants $\lambda'_{ijk}$ are present.
Still, the ${\tilde c}_L$ interpretation of the HERA data 
suggests that the R-parity conserving modes have moderate
rates and that
${\cal B}(K^+\to \pi^+ \nu \bar \nu )$ is very close
to the current experimental limits.

The second valence production mechanism is
$e^+ d \rightarrow {\tilde t}_L$ via $\lambda'_{131}$.
For this coupling, the larger constraint
arises from atomic parity violation~\cite{noi}.
The most recent value for the bound is
$\vert \lambda'_{131} \vert < 0.08
\left({m_{{\tilde t}_L} / 200 
~\hbox{GeV}}\right)$\cite{atpar1,atpar2}, allowing 
for a sufficient production rate,
for ${\cal B}$ $\geq 0.25$.

On the other hand,
sea production processes are excluded by a combination
of different experimental constraints;
${\tilde u}_L$ production from sea quarks of
the second or third generation is constrained by
limits on charged-current
universality. Moreover,
contributions to the electron neutrino mass,
rule out sea-quark production mechanisms involving only second or
third generation particles. 
Finally, the sea process $e^+ \bar u \to \bar{\tilde{d}}_{k_R}$
is also excluded, since otherwise
a much larger effect would have been observed in
$e^-  u \to {\tilde{d}}_{k_R}$.
Then, the only sea production mechanism that survives
is $e^+s\to {\tilde t}_L$ via the $\lambda'_{132}$ 
coupling~\cite{AEGLM}.
This has been examined in detail, including
stop mixing effects~\cite{ELS}.

Let us now pass to the
 squark decay modes. ${\tilde c}_L$, 
has the following decays:
${\tilde c}_L\rightarrow c \chi^0_i$,
${\tilde c}_L\rightarrow s \chi^+_j$ and
${\tilde c}_L\rightarrow d e^+$,
where $\chi^0_i, \chi^+_j$ are neutralinos and charginos.
The decay rate for the $R$-parity violating mode 
(in the absence of stop mixing) is 
\bea
\Gamma  = \frac{1}{16 \pi} (\lambda'_{121})^2 
m_{{\tilde c}_L} \nonumber
\eea
The R-conserving decays are suppressed either by phase space
or, in the ${\tilde c}_L\to c \chi^0_i$ case, 
by cancellations in the neutralino
couplings~\cite{AEGLM}.
The decay rate of $\tilde{c}_L$ to neutralinos is
\bea
\Gamma' = 
\frac{g^2}{32 \pi} m_{\tilde{c}_L} \left \{
\left ( \frac{ m_c N_{i4}}{ M_W \sin\beta}\right)^2+
\left (N_{i2}+\frac{1}{3}\tan\theta_{W} N_{i1}
\right ) ^2 \right\}
\left
(1-\frac{m^2_{\chi^0_i}}{m_{\tilde{c}_L}^2}\right )^2 \nonumber
\eea
Here $N_{ij}$ are the elements of the unitary matrix
that diagonalises the neutralino mass matrix
in the $SU(2) - U(1)$ gaugino basis and $\tan \beta$ is the
ratio of Higgs vacuum expectation values. 
For $\tilde{c}_L$ decays, the term
proportional to $m_c$ can be neglected, while
the second term is very small,
either in the higgsino region
or if there is a cancellation
$N_{12} \sim - {1 \over 3} \hbox{tan} \theta_W N_{11}$,
for the lighter neutralino~\cite{AEGLM}.
This cancellation
occurs in an acceptable domain of the
supersymmetric parameter space,
as we can see from Fig.1a. 
Here, the region where charginos have mass
less than 85 GeV, has been subtracted from the plot
(for neutralinos, the LEP2 bounds are much weaker).
In the figure,
$M_2$ is the $SU(2)$ gaugino mass, while the $U(1)$ gaugino mass 
is determined by the unification 
relation $M_1=(5/3)\tan^2\theta_WM_2$.

It is interesting to note that such a cancellation is
not present in the coupling of the lightest
neutralino to other sfermions. One type of example 
is given by the couplings to the
$SU(2)$ singlets $\tilde{u}_R$,
$\tilde{d}_R$ and $\tilde{e}_R$.
A second example
is the coupling of the neutralino to
$\tilde{\nu}_L$, which is of interest, since
$\tilde{\nu}_L$ could in principle be produced
at LEP2, provided any of the $L_1L_{2,3}\bar{E}_1$ 
operators is sufficiently large \cite{noi,DL,kal2}.
As shown in Fig.1b, there is no analogous
effect as in the $\tilde{c}_L$ case and 
the decay channels are determined by 
the mass of the squark and the gauginos.

If we drop the unification condition for gaugino masses,
it is possible to go to examples where charginos
can be quite heavy, while neutralinos are light. This
would occur for large $M_2$, but small $M_1$ values.
In such a case, only bounds on neutralino masses from
LEP2 would be relevant.  For completeness, we show an 
indicative plot in Fig.1c.

In the case of ${\tilde t}_L$, the
neutralino decay mode ${\tilde t}_L \to t 
\chi^0_i$ is kinematically closed 
in a natural way
and large values of $\cal B$ are 
obtained for the region where the chargino decays
are suppressed by phase-space.
We give the contour
plots for this case in Fig.1d. In this particular
figure, instead of fixing $\lambda$, we use 
$\lambda/\sqrt{\cal B}$. This is done
in order to see more explicitly
which is the region of parameter space where both
the R-violating and the R-conserving decay modes
are non-negligible.

Possible tests  towards checking the
various schemes, include $e^-$ 
and polarised beam runs at HERA, as well as 
search for the cascade
decays that result from the R-conserving vertices.
Squark pair production or single slepton production at
the Tevatron, virtual effects at LEP2,
observable contributions
to $K \rightarrow \pi \nu \bar{\nu}$
and/or neutrinoless $\beta\beta$ decays, are also
among the possibilities to consider.

\begin{figure}
\centerline{\epsfig{figure=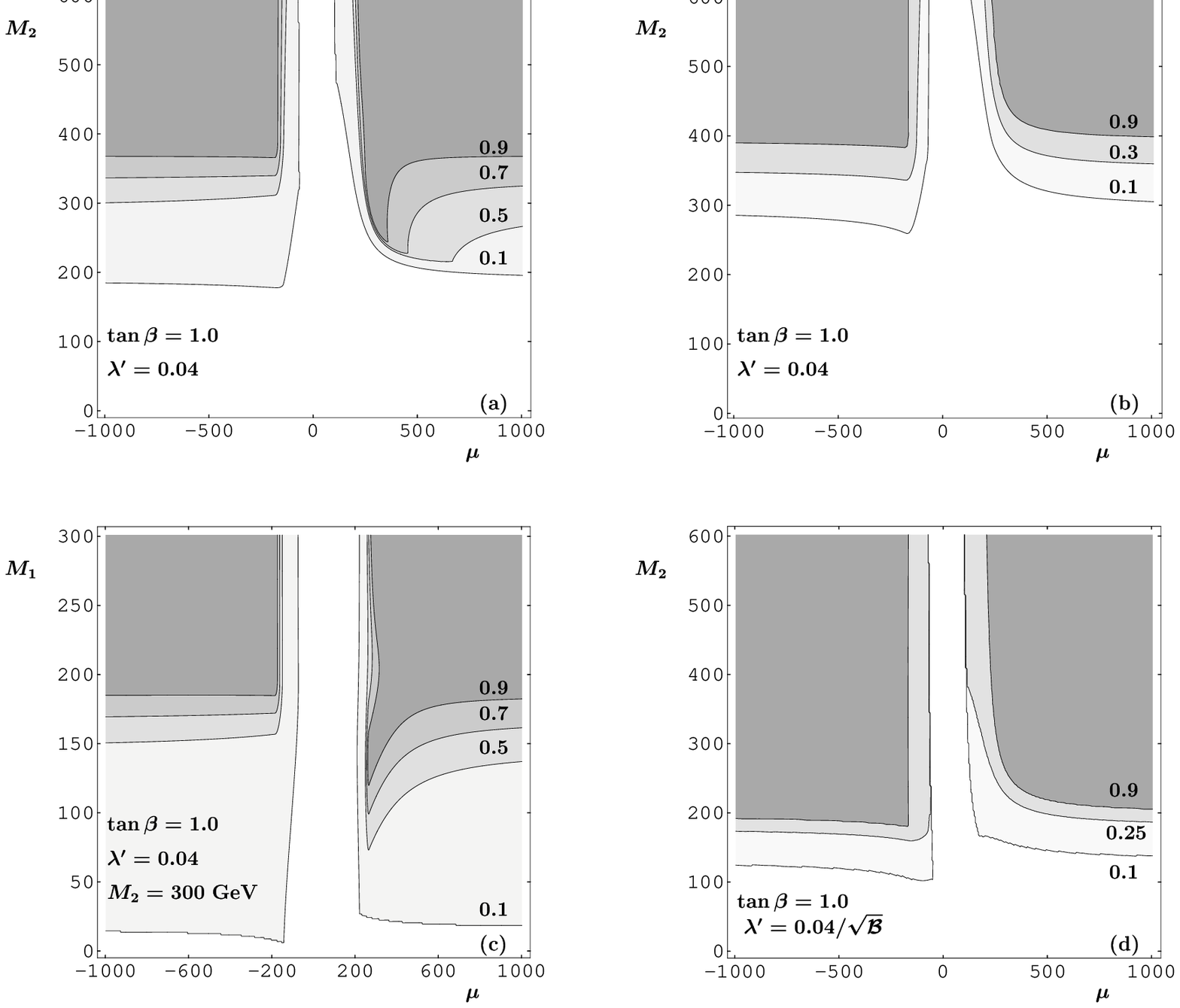,width=1.13\textwidth}}
\vspace*{-9.0 cm}
\ccaption{}{ \label{fig:contours}    
Contours of ${\cal B}(e d)$ for the $R$-violating decays, 
in the $\mu-M_2$ plane. Fig.1a
is for $\tilde{c}_L$ decays, 
imposing the unification
relation for gaugino masses. Fig.1b is the
respective figure for
$\tilde{\nu}_L$ decays.
Fig.1c is an example of
$\tilde{c}_L$ decays, and
arbitrary $M_1$ and $M_2$.
Finally Fig.1d shows the contours for
$\tilde{t}_L$ decays 
and $M_1=(5/3)\tan^2\theta_WM_2$.
$\lambda^{\prime}$ has been fixed to 0.04, except
for the last figure, 
where $\lambda' = 0.04/\sqrt{{\cal B}}$.
In all plots, $tan\beta=1$ and $m_{\tilde{q}}=200$ GeV. 
The region with a light chargino mass $ <$ 85 GeV 
has been subtracted in (a),(b) and (d),
while in (c) charginos are heavy and
we applied the LEP2 bounds on neutralinos instead.}
\end{figure}

{\tt {\bf Acknowledgements:}}
I would like to thank G. Altarelli, J. Ellis,
G.F. Giudice and M.L. Mangano for a very enjoyable 
collaboration, leading to the presented results.
My work is funded by a
Marie Curie Fellowship (TMR-ERBFMBICT-950565).
Financial support from CERN and the DIS97 organising
committee, for attending
the conference, is gratefully acknowledged.

\pagebreak

\end{document}